\documentclass[prb,floatfix,superscriptaddress,amsmath,amssymb,twocolumn,showpacs]{revtex4}
\usepackage{pstricks}
\usepackage[latin1]{inputenc}
\usepackage{amsmath}
\usepackage{amssymb}
\usepackage{graphicx}
\usepackage{graphics}
\usepackage{color}
\usepackage{natbib}

\def \ba {\begin{eqnarray}}
\def \ea {\end{eqnarray}}

\begin{document}

\title{Diagnosing order by disorder in quantum spin systems}

\author{C. A.\ Lamas}
\affiliation{IFLP-CONICET. Departamento de F\'isica, Universidad Nacional de La Plata,
C.C.\ 67, 1900 La Plata, Argentina.}

\author{D. C. Cabra}
\affiliation{IFLP-CONICET. Departamento de F\'isica, Universidad Nacional de La Plata,
C.C.\ 67, 1900 La Plata, Argentina.}

\author{P.\ Pujol}
\affiliation{Laboratoire de Physique Th\'eorique,  IRSAMC, CNRS and Universit\'e de Toulouse, UPS, F-31062 Toulouse, France}

\author{G. L. Rossini}
\affiliation{IFLP-CONICET. Departamento de F\'isica, Universidad Nacional de La Plata,
C.C.\ 67, 1900 La Plata, Argentina.}

\date{\today}


\begin{abstract}
In this paper we study the frustrated $J_{1}-J_{2}$ quantum Heisenberg model
on the square lattice for $J_2 > 2 J_1$, in a magnetic field. In this regime  the classical system is known to have a degenerate
manifold of lowest energy configurations, where standard thermal order by disorder occurs.
In order to study its quantum version we use
a path integral formulation in terms of coherent states.
We show that the classical degeneracy in the plane transverse to the magnetic field is lifted by  quantum fluctuations. Collinear states are then selected, in a similar pattern to that set by thermal order by disorder, leaving a $Z_2$ degeneracy.
A careful analysis reveals a purely quantum mechanical effect given by the tunneling between the two minima selected by fluctuations.
The effective description contains two planar ($XY$-like) fields conjugate to the total magnetization and the difference of the two sublattice magnetizations.
Disorder in either or both of these fields produces the locking of their conjugate observables.
Furthermore, within this scenario we argue that the quantum state is close to a product state.

\end{abstract}

\maketitle

\section{Introduction}
\label{sec:Introduction}
The properties of the two-dimensional (2D) Heisenberg model have received considerable interest in the last
years\cite{Henley,Villain_77,Loison-Simon_2000,Cabra-Lamas_2011,Capriotti_2004,Lamas_2009,Carretta-Melzi_2002,Melzi-Carretta_2000,Zhang-Arlego_2014,Zhang-Lamas_2013,Oshikawa_1999,TTH,Lamas-Ralko_2012,Lamas-Ralko_2013,Zhitomirsky-Honecker_2000},
in part because of the possible connection between magnetism and high-temperature superconductivity.
In this sense, one of the most typical examples of a two-dimensional frustrated spin system is given by the $J_{1}-J_{2}$ antiferromagnetic Heisenberg model on the square-lattice.

The classical version of this model
has a continuous manifold of degenerate ground states related by the rotation
of one sub-lattice with respect to the other.
Thermal and quantum fluctuations can stabilize collinear spin configurations \cite{Henley},
a particular case of the phenomenon known as Order By Disorder (OBD) \cite{Villain_77}.
It is generally accepted that quantum and thermal fluctuation select the
same ground state from the classical manifold. However,
there exists some examples where the quantum fluctuations select a different ground
state\cite{honecker_kagome1/3} than thermal ones.
%
For this reason, it is interesting to study the quantum and thermal
contribution to the free-energy in order to distinguish whether thermal and quantum selections may be different. 

Besides its theoretical interest, this model is also important because of the existence of
compounds like Li$_2$VOSiO$_4$, Li$_2$VOGeO$_4$ and VOMoO$_4$ in which dominant magnetic interactions
consist in first and second nearest neighbors exchange \cite{Melzi-Carretta_2000,Carretta-Melzi_2002}.

The path integral description of the magnetic degrees of freedom in terms of coherent states represents a good alternative to
study ordering due to disorder phenomena in partially polarized spin systems. 
Under a homogeneous magnetic field, a semiclassical approach\cite{TTH} leads to planar $XY$-like fields describing the spin components in the transverse plane, 
canonically conjugate to magnetization degrees of freedom along the external field. 
The presence of topological terms (Berry phases) with coefficients that depend on the total magnetization may either allow or forbid the vortex proliferation that disorders the planar degrees of freedom. Delocalization of the spin components in the transverse plane is then related to the total magnetization.
This has allowed for the study of 
plateaux formation in magnetization curves\cite{Lamas-Pujol_2010}.

Topological terms are also relevant in the quantum order by disorder selection, 
controlling the weight of field configurations with non-vanishing vorticity. 
For the strongly frustrated $J_{1}-J_{2}$ antiferromagnetic Heisenberg model on the square lattice, in the presence of an external magnetic field,
the effective description contains two planar $XY$-like fields. 
One of them (hereafter the symmetric field) is conjugate to the total magnetization along the magnetic field, while the other (the antisymmetric field) is conjugate to the difference between bipartite sublattice magnetizations.
For the latter, an effective potential is found in the action allowing for possible non-trivial instanton-like excitation processes. 
By tuning the couplings it may be possible to reduce the spin stiffness and favor the proliferation of such instanton configurations until the system enters into a disordered phase where broken symmetries are recovered.

\section{The Model and its Classical degeneracy}
\label{sec:Model}

We study a spin-S next-nearest-neighbor antiferromagnetic Heisenberg model on the square lattice in the presence
of a homogeneous magnetic field. The Hamiltonian is given by
\ba
\nonumber
H&=&\displaystyle{J_{1}\sum_{NN} \vec{S}(\vec{r})\cdot \vec{S}(\vec{r}')}
+\displaystyle{J_{2}\sum_{NNN} \vec{S}(\vec{r})\cdot \vec{S}(\vec{r}')}\\\label{eq:Hamiltonian}
&-&h\sum_{\vec{r}}S^{z}(\vec{r})
\ea
where $J_{1}$, $J_{2}$ are positive. The magnetic field $h$ points in the $z$ direction,
vectors $\vec{r}$ and $\vec{r}'$ belong to the two dimensional square lattice $\vec{r}=n_{x}\hat{x}+n_{y}\hat{y}$,
with $\hat{x}=a(1,0)$, $\hat{y}=a(0,1)$,
$a$ is the lattice spacing  and the summations denoted as $NN$ and $NNN$ run on nearest-neighbor and next-nearest-neighbor sites respectively.
%
Despite its simplicity, this model is paradigmatic in quantum magnetism since it shows order from disorder selection, a magnetization plateau at $M=1/2$,
and field induced ordering\cite{Zhitomirsky-Honecker_2000}.
In the following sections we study the degenerate classical ground state and the low energy theory describing thermal and quantum fluctuations around it.
%

%

We start by parameterizing the ground state manifold as
\ba
\nonumber
\vec{S}_{l}(\vec{r})=\left(
\begin{array}{c}
 \sqrt{S^{2}-m^{2}}\cos(\vec{Q}\cdot\vec{r}+\alpha_l)\\
\sqrt{S^{2}-m^{2}}\sin(\vec{Q}\cdot\vec{r}+\alpha_l)\\
m
\end{array}
 \right),
\ea
where $l=1,2$ is a sub-lattice index, $Q$ is the pitch angle and $\alpha_l$ is a sub-lattice dependent phase.

At zero magnetic field  and $J_2<2J_1$, the minimum energy configuration corresponds to a N\'eel order whereas for $J_2>2J_1$
the ground state breaks up into two square $\sqrt{2} \times \sqrt{2}$ sub-lattices\cite{Henley}.
Each sub-lattice is ordered antiferromagnetically, leaving  a classical ground state degeneracy
associated to global rotations of all the spins belonging to one of the sub-lattices.
To label the degenerate ground states we can choose one reference spin from each sub-lattice, and use the relative
angle $\alpha=\alpha_2-\alpha_1$ between these two spins to parameterize the non-trivial degeneracy,
the ground-state energy being independent of $\alpha$.
In order to represent the ground state manifold, we take in the following
 $\vec{Q}=(\frac{\pi}{a},0)$, $\alpha_1=0$ and $\alpha_2=\alpha$.

\begin{figure}[t!]
\includegraphics[width=0.42\textwidth]{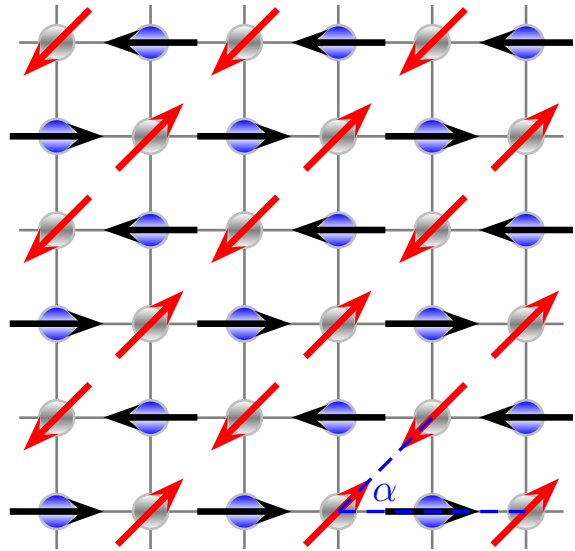}
\caption{(Color on-line) Sketch of a given classical configuration on the plane. Each sub-lattice is ordered antiferromagnetically with a relative
 angle $\alpha$ between sub-lattices.}
\label{fig:classical_gs}
\end{figure}

In the presence of a magnetic field the classical spins are canted towards the field direction
 and the classical energy can be written as
\ba
\nonumber
\frac{E}{N_{c}}=4m^{2}J_{1}+J_{2}(8m^{2}-4S^{2})-2mh,
\ea
where $N_{c}$ is the number of unit cells. Minimizing with respect to $m$ we obtain
%
\ba
\nonumber
m=\frac{h}{(4J_{1}+8J_{2})}.
\ea
In Fig. \ref{fig:classical_gs} we show the projection on the $x-y$ plane of one of the many configurations that minimize the
energy of the system.

\section{Low energy effective action.}

At non-zero magnetic field, since we are dealing with a magnetized classical state we can use a particular path integral approach in terms of coherent states\cite{TTH,Lamas-Pujol_2010} to estimate the free energy of the system.
First, we write two fields by sub-lattice to represent the spins as
\ba
\nonumber
\vec{S}_{l}(\vec{r})=\left(
\begin{array}{c}
 S\sin(\theta_{l}(\vec{r})\cos(\phi_{l}(\vec{r}))\\
S\sin(\theta_{l}(\vec{r}))\sin(\phi_{l}(\vec{r}))\\
S\cos(\theta_{l}(\vec{r}))
\end{array}
 \right)
\ea
and parameterize the fluctuations around the classical solution in terms of two fields $\varphi$ and $\delta \theta$,
\ba
\nonumber
\phi_{l}(\vec{r})&=&\vec{Q}\cdot \vec{r}+\alpha_{l}+\varphi_{l}(\vec{r})\\\nonumber
\theta_{l}(\vec{r})&=&\theta_{0}+\delta\theta_{l}(\vec{r}).
\ea
where $\theta_{0}$ is the classical solution ($m=S\cos{\theta_{0}}$), which in the present case
is given by $\theta_{0}=\arccos\left(\frac{h}{S(4J_{1}+8J_{2})}\right)$.
As a first step towards a Hamiltonian theory for spin fluctuations, one takes $\varphi$ and $\delta \theta$ as canonical conjugates\cite{TTH}.

Expanding up to second order in the fluctuating fields we have
\ba
\nonumber
S\cos{(\theta_{0}+\delta \theta_{l}(\vec{r}))}&\simeq&  S\cos{\theta_{0}}-
S \sin{\theta_{0}}\; \delta \theta_{l}(\vec{r})\\\nonumber
&-&\frac{S}{2}\cos{\theta_{0}}(\delta \theta_{l}(\vec{r}))^{2}\\\nonumber
S\sin{(\theta_{0}+\delta \theta_{l}(\vec{r}))}&\simeq&  S\sin{\theta_{0}}+
S \cos{\theta_{0}}\; \delta \theta_{l}(\vec{r})\\\nonumber
&-&\frac{S}{2}\sin{\theta_{0}}(\delta \theta_{l}(\vec{r}))^{2}.
\ea
If we calculate the Poisson brackets $\{ S^{z},S^{\pm}\}_{\varphi,\delta\theta}$, we obtain
 $i \hbar\{ S^{z},S^{\pm}\}_{\varphi,\delta\theta}=-S(\sin{\theta_{0}}
-\delta\theta\cos{\theta_{0}})\left( \pm \hbar S^{\pm}  \right)$.
In order to generate the correct $SU(2)$ algebra, one redefines the canonical conjugate for $\varphi_{l}$ as 
\ba
\frac{a^{2}}{S}\Pi_{l}(\vec{r})=-\sin{\theta_{0}}\delta \theta_{l}(\vec{r})
-\frac{1}{2}\cos{\theta_{0}}(\delta \theta_{l}(\vec{r}))^{2}
\ea
%
The spin operators in terms of the new conjugate pairs read
\ba
\nonumber
S_{l}^{\pm}(\vec{r})&=&e^{\pm i \vec{Q}\cdot \vec{r}}e^{\pm i \varphi_{l}(\vec{r})}
\left[
S\sin{\theta_{0}}  -a^{2}\frac{m}{S\sin{\theta_{0}}}\Pi_{l}(\vec{r})\right.\\\label{eq:Spin_pm}
&-&\left. a^{4}\frac{S^{2}}{S^{2}-m^{2}}\frac{1}{S\sin{\theta_{0}}}\Pi_{l}^{2}(\vec{r})
\right],\\\label{eq:spin_Sz}
S_{l}^{z}(\vec{r})&=&m + a^{2}\Pi_{l}(\vec{r}).
\ea
Notice from the last equation that the $\Pi_l$ fields describe fluctuations in the spin components along the magnetic field.

Using expressions (\ref{eq:Spin_pm}) and (\ref{eq:spin_Sz}) in Hamiltonian (\ref{eq:Hamiltonian}), taking the continuum limit
and retaining terms up to second order in the fields, 
we can write $H=H_{\phi}+H_{\pi}+H_{0}$.
Here $H_{0}$ is a contribution independent of the fields 
while $H_{\phi}$ and $H_{\pi}$ are given by
\small
\ba
\nonumber
H_{\varphi}\!\!&=&\!\!\int d^{2}r
\left\{  \phantom{\frac12} \!\!\!\!\!\! \right.
\left.
(S^2\!-\!m^2)(J_{2}+\frac{J_{1}}{2}\cos(\alpha))[(\partial_{x}\varphi_{1})^2+(\partial_{x}\varphi_{2})^2]
\right.\\\nonumber
&+& \left.
(S^2-m^2)(J_{2}-\frac{J_{1}}{2}\cos(\alpha))[(\partial_{y}\varphi_{1})^2+(\partial_{y}\varphi_{2})^2]
\;\; \right\}\\\nonumber
H_{\pi}&=&\int d^{2}r
\left\{  \phantom{\frac12} \!\!\!\!
4 J_{1} a^2\Pi_{1}\Pi_{2}
+4 J_{2} a^2\left(\Pi_{1}^2 + \Pi_{2}^2 \right)
\right\}.
\ea
\normalsize
%
The  $\vec{r}$ dependendence in the fields has been omitted for
simplicity.
As a further step towards a path-integral formulation, we write an effective action as
%
%

\ba
\nonumber
\mathcal{S} = \mathcal{S}_{\varphi} + \mathcal{S}_{\pi} \label{Total_S},
\ea
where
\small
\ba
\nonumber
\mathcal{S}_{\varphi}\!\!&=&\!\!\int d^{2}r \int d\tau
\left\{  \phantom{\frac12} \!\!\!\!\!\!\!\! \right.
\left.
(S^2-m^2)(J_{2}+\frac{J_{1}}{2}\cos(\alpha))[(\partial_{x}\varphi_{1})^2+(\partial_{x}\varphi_{2})^2]
\right.\\\nonumber
&+& \left.
(S^2-m^2)(J_{2}-\frac{J_{1}}{2}\cos(\alpha))[(\partial_{y}\varphi_{1})^2+(\partial_{y}\varphi_{2})^2]\right.\\\nonumber
&+& \left. i\left(\frac{S-m}{a^2}\right) \left(\partial_{\tau}\varphi_{1}+\partial_{\tau}\varphi_{2}\right)
\;\; \right\}\\\nonumber
\mathcal{S}_{\pi}\!\!&=&\!\!\int d^{2}r \int d\tau
\left\{  \phantom{\frac12} \!\!\!\!
4 J_{1} a^2\Pi_{1}\Pi_{2}
+4 J_{2} a^2\left(\Pi_{1}^2 + \Pi_{2}^2 \right)\right. \\
&-&\left.  i \Pi_{1}\partial_{\tau}\varphi_{1}-i \Pi_{2}\partial_{\tau}\varphi_{2}
\phantom{\frac12}\!\!\!\!\right\}.
\ea
\normalsize

\begin{figure}[t!]
\includegraphics[width=0.45\textwidth]{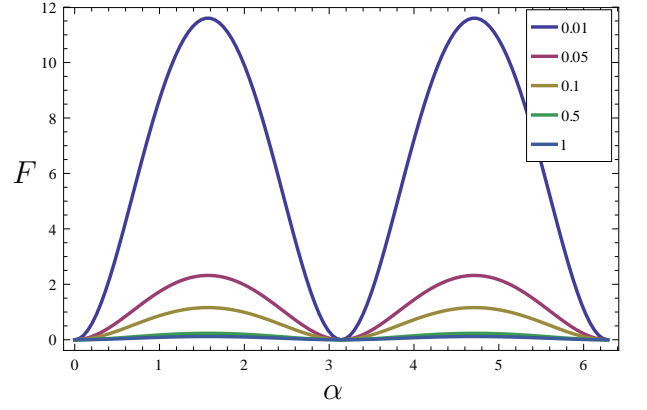}
\caption{(Color on-line) Free energy in terms of the sub-lattices relative angle $\alpha$.
The plot corresponds to $J_{2}=0.6 J_1$ and low magnetization, for different values of $\beta$
denoted in the plot legend.
Minimum values correspond to collinear configurations ($\alpha=0$ and $\alpha=\pi$).}
\label{fig:numerical_Z_h0}
\end{figure}
%
%

After writing the path-integral one can readily integrate out the Gaussian $\Pi$-fields to get the effective action:

%
\small
\ba
\nonumber
\mathcal{S}_{eff} \!\! &=&\!\! \int d^{2}r \int d\tau
\left\{  \phantom{\frac12} \!\!\!\!\!\!  \right.
\left.
(S^2\!-\!m^2)(J_{2}+\frac{J_{1}}{2}\cos(\alpha))[(\partial_{x}\varphi_{1})^2\!+\!(\partial_{x}\varphi_{2})^2]
\right.\\\nonumber
&+& \left.
(S^2-m^2)(J_{2}-\frac{J_{1}}{2}\cos(\alpha))[(\partial_{y}\varphi_{1})^2+(\partial_{y}\varphi_{2})^2]\right.\\\nonumber
&+&  \frac{1}{(\Gamma_2+\Gamma_1)}\left(
\partial_{\tau}\varphi_{1}+\partial_{\tau}\varphi_{2}
\right)^{2}
+\frac{1}{(\Gamma_2-\Gamma_1)}\left(
\partial_{\tau}\varphi_{1}-\partial_{\tau}\varphi_{2}
\right)^{2}\\
&+& \left. i\left(\frac{S-m}{a^2}\right) \left(\partial_{\tau}\varphi_{1}
+\partial_{\tau}\varphi_{2}\right)
\;\; \right\}
\label{Total_action}
\ea
\normalsize
where $\Gamma_2=8 J_{2} a^2 $ and $\Gamma_1= 4 J_{1}a^{2}$.
Notice that one can write this action in a decoupled form in terms of symmetrical and anti-symmetrical fields
\ba
\phi_{s}=\frac12(\varphi_{1}+\varphi_{2}),\\
\phi_{a}=\frac12(\varphi_{1}-\varphi_{2}),
\ea
which yields
\ba
\nonumber
\mathcal{S}_{eff} =\mathcal{S}_{s}+\mathcal{S}_{a},
\ea
with
\small
\ba
\nonumber
\mathcal{S}_s\!\! &=&\!\! \int d^{2}r \int d\tau
\left\{  \phantom{\frac12} \!\!\!\!\!\!  \right.
\left.
2(S^2\!-\!m^2)(J_{2}+\frac{J_{1}}{2}\cos(\alpha))(\partial_{x}\phi_{s})^2
\right.\\\nonumber
&+& \left.
2(S^2-m^2)(J_{2}-\frac{J_{1}}{2}\cos(\alpha))(\partial_{y}\phi_{s})^2
\right.\\
&+&  \frac{4}{(\Gamma_2+\Gamma_1)}\left(
\partial_{\tau}\phi_{s}\right)^{2}
+ \left. i2\left(\frac{S-m}{a^2}\right)
\left(\partial_{\tau}\phi_{s}\right)
\;\; \right\}
\label{eq:sym_action}
\ea
\normalsize
\small
\ba
\nonumber
\mathcal{S}_{a}\!\! &=&\!\! \int d^{2}r \int d\tau
\left\{  \phantom{\frac12} \!\!\!\!\!\!  \right.
\left.
2(S^2\!-\!m^2)(J_{2}+\frac{J_{1}}{2}\cos(\alpha))(\partial_{x}\phi_{a})^2
\right.\\\nonumber
&+& \left.
2(S^2-m^2)(J_{2}-\frac{J_{1}}{2}\cos(\alpha))(\partial_{y}\phi_{a})^2
\right.\\
&+&  \frac{4}{(\Gamma_2-\Gamma_1)}\left(
\partial_{\tau}\phi_{a}\right)^{2}
 \left. \;\; \right\}
\label{eq:anti-sym_action}
\ea
\normalsize

On generic grounds, the issue about the presence of magnetization plateaux is closely related to the value of the Berry phase term in the Eq.(\ref{eq:sym_action}) (see for instance [\onlinecite{TTH}]).
The canonical conjugates to the symmetric and anti-symmetric fields are  $\Pi_1+ \Pi_2$ and  $\Pi_1- \Pi_2$. The first one is related to the total magnetization and the second controls the difference of magnetization between sub-lattices, also known as spin imbalance.

\section{Analysis of the vortex free regime}
\label{sec:Free_energy}

Let us consider the preceding action  first assuming that the fields $\varphi_{\alpha}$ ($\alpha = 1,~2$)
have only vortex free configurations.
These regular fields can be treated as standard periodic scalar fields: one can Fourier transform and express the action in terms of the momentum and Matsubara frequency variables
\ba
\nonumber
\varphi_{\alpha}(\vec{r},\tau)=\sum_{n=-\infty}^{\infty}\frac{1}{(2\pi)\beta}\int d^2 k
e^{i\vec{k}\cdot\vec{r}} e^{-i\omega_{n}\tau}\varphi_{\alpha}(\vec{k},\omega_{n})
\ea
where $\omega_{n}=\frac{2\pi n}{\beta}$.
We obtain for the action
\small
\begin{widetext}
\ba
\nonumber
\mathcal{S}=
\frac{1}{(2\pi)^2}\!\!\sum_{n=-\infty}^{\infty} \! \int d^2 k
\left\{
(\epsilon^2 \!\! + \!\! \frac{2\Gamma_{2}}{\Gamma_{2}^2-\Gamma_{1}^2} \omega_{n}^2)(|\varphi_{1}(\vec{k},\omega_{n})|^2 \!\! +\!\! |\varphi_{2}(\vec{k},\omega_{n})|^2)\right.
\!-\!\left. (\frac{2\Gamma_1}{\Gamma_{2}^2-\Gamma_{1}^2} \omega_{n}^2 )
(\varphi_{1}(\vec{k},\omega_{n})\varphi^{*}_{2}(\vec{k},\omega_{n})\!+\!\varphi^{*}_{1}(\vec{k},\omega_{n})\varphi_{2}(\vec{k},\omega_{n}))
\right\}
\ea
\end{widetext}
\normalsize
where
\small
\ba
\!\!\epsilon^2\!=\!(S^2\!-\!m^2)\!\left[(J_{2}\!+\!\frac{J_{1}}{2}\cos(\alpha))k_{x}^2\!+\!(J_{2}\!-\!\frac{J_{1}}{2}\cos(\alpha))k_{y}^2\right]
\ea
\normalsize
We can evaluate $\mathcal{Z}=\int \mathcal{D}[\phi] e^{-\mathcal{S}}$ by integrating the fields,
\small
\ba
\nonumber
\log(\mathcal{Z})=N(\beta)&-&\frac12\sum_{n=-\infty}^{\infty} \int \frac{d^2 k}{(2\pi)^2}
\log\left[ (\epsilon^2+   \frac{2\Gamma_{2}}{\Gamma_{2}^2-\Gamma_{1}^2} \omega_{n}^2)^2\right.\\
&-&\left. (\frac{2\Gamma_1}{\Gamma_{2}^2-\Gamma_{1}^2} \omega_{n}^2 )^2\right].
\ea
\normalsize
Notice that all dependence in $\vec{k}$ and the parameter $\alpha$ is contained in $\epsilon$. After some algebra
we can write
\ba
\nonumber
\log(\mathcal{Z})=N'(\beta)&-&\frac12\sum_{n=-\infty}^{\infty} \int \frac{d^2 k}{(2\pi)^2}
\log\left[\omega_{n}^4\right. \\
&+&\left. x \epsilon^2 \omega_{n}^4  \frac14 (\Gamma_{2}^2-\Gamma_{1}^2)\epsilon^4\right]\\\nonumber
=N'(\beta)&-&\frac12\sum_{n=-\infty}^{\infty} \int \frac{d^2 k}{(2\pi)^2}\big\{
\log\left[\omega_{n}^{2}+\omega_{+}^{2}\right]+\\
&+&\log\left[\omega_{n}^{2}+\omega_{-}^{2}\right]
\big\}
\ea
where $\omega_{\pm}=\frac{\epsilon}{\sqrt{2}}\sqrt{\Gamma_{2}\pm \Gamma_{1}}$.
%
Now we use that
\ba
\log(\omega_{n}^{2}+\omega_{\pm}^{2})=\log(\omega_{n}^2+\frac{1}{\beta^2}) + \int_{\frac{1}{\beta^2}}^{\omega_{\pm}^{2}} \frac{d(t^{2})}{\omega_{n}^2+t^2}
\ea
and $\log(\omega_{n}^2+\frac{1}{\beta^2})=-2\log(\beta)+\log(2\pi^2 n^2 +1)$.
Then we have
\small
\ba
\log(\mathcal{Z})&=&
\tilde{N}(\beta)-\frac12\sum_{n=-\infty}^{\infty} \int \frac{d^2 k}{(2\pi)^2}\left\{
\int_{\frac{1}{\beta^2}}^{\omega_{+}^{2}} \frac{d(t^{2})}{\omega_{n}^2+t^2}\right.\\\nonumber
&+&\left.\int_{\frac{1}{\beta^2}}^{\omega_{-}^{2}} \frac{d(u^{2})}{\omega_{n}^2+u^2}
\right\},
\ea
\normalsize
where we have included all the  vacuum contributions in the first term. Now we can perform the summation to obtain
\small
\ba
\nonumber
\log(\mathcal{Z})&=&\int \frac{d^2 k}{(2\pi)^2}\big\{
\log\left[csch\left(\frac{\beta\omega_{+}}{2}\right)\right]\\\nonumber
&+&\log\left[csch\left(\frac{\beta\omega_{-}}{2}\right)\right]
\big\}.
\ea
\normalsize
Finally, after some rearrangements, we obtain an expression for the free energy, 
%
%
\ba
\label{eq:free_energy_general}
F&=&\frac12 \int \frac{d^2 k}{(2\pi)^2}
(\omega_{+}+\omega_{-})\\\nonumber
&+&\frac{1}{\beta}\int \frac{d^2 k}{(2\pi)^2} \left[\log\left(1-e^{-\beta\omega_{+}}\right)+\log\left(1-e^{-\beta\omega_{-}}\right)\right].
\ea

Equation (\ref{eq:free_energy_general}) makes apparent the role of quantum and thermal fluctuations: the first term represents the quantum zero point contribution to the free energy whereas the second term is the thermal contribution.
We show in figure (\ref{fig:numerical_Z_h0}) plots of the free energy in terms of the relative angle between sub-lattices, for low magnetization and several temperatures. The free energy shows (as expected) two minima.
In the present model the minima of both quantum and thermal contributions are located in $\alpha=0$ and $\alpha=\pi$, corresponding to the collinear configurations $(0,\pi)$ and $(\pi,0)$.
This translates into an emergent $Z_{2}$ symmetry of the system at large scales. The selection of an angle, at either the values $0$ or $\pi$ implies the spontaneous breaking of the $Z_{2}$ symmetry.
One expects this symmetry to be restored at higher temperatures.
%

%

\section{Static approximation and classical limit}
From equation (\ref{eq:free_energy_general}) it is easy to extract the zero temperature  (quantum) contribution $F_{Q}$, which is simply the first term, and the thermal contribution of the purely classical model $F_{Cl}$. The latter is obtained from the second term in the limit $\beta \to 0$ whose dominant contribution is given, up to $\alpha$ independent terms, by:
$$
F_{Cl} = \frac{1}{\beta}\int \frac{d^2 k}{(2\pi)^2} \log\left[\omega_{+} \omega_{-} \right] ~.
$$
This result could have been obtained simply by implementing the static approximation in eq.
(\ref{Total_S}) or even (\ref{Total_action}) which is obtained by equating to zero all the time derivatives and making the replacement:
$$
\int d\tau \to \beta ~.
$$
Both $F_{Cl}$ and $F_{Q}$ coincide with the results obtained by Henley \cite{Henley} for the planar model using classical and quantum spin wave theory. Our approach provides nevertheless a more complete analysis,
allowing us to study the crossover from the quantum to the purely classical regime and more importantly, allowing the study of non-trivial topological contributions given by the Berry phase terms.

\section{Magnetization dependence of the $Z_{2}$ and KT transition temperature in the classical regime}
An emergent $Z_{2}$ chirality symmetry in a continuous frustrated magnet can be found in many examples as the $J_1-J_2$ $XY$
model \cite{Henley}, the $J_1-J_2$ Heisenberg model as well as the fully frustrated $XY$ model \cite{Villain_77}.
At low temperatures one expects a quasi-long-range order for the spin variables and an ordered pattern  for the chiral degree of freedom (broken $Z_{2}$ symmetry) while at high temperatures both degrees of freedom should be disordered with short range correlations.
One then expects a Kosterlitz-Thouless (KT)  transition at finite temperature $T_{KT}$ for the spin degrees of freedom and an Ising-like transition
for the chiral degrees of freedom at a temperature $T_{Ising} \geq T_{KT}$ (for restoring $Z_2$ symmetry implies disorder). 
Numerical Monte Carlo results \cite{Loison-Simon_2000} suggest that these critical temperatures, although very close, are different.
Moreover, the model seems to have a rather large cross-over scale at the vicinity of the critical points making quite difficult
the confirmation that the nature of the transitions are indeed of the traditional Ising and KT type \cite{Hasenbusch,KT}.

\begin{figure}[t!]
\includegraphics[width=0.46\textwidth]{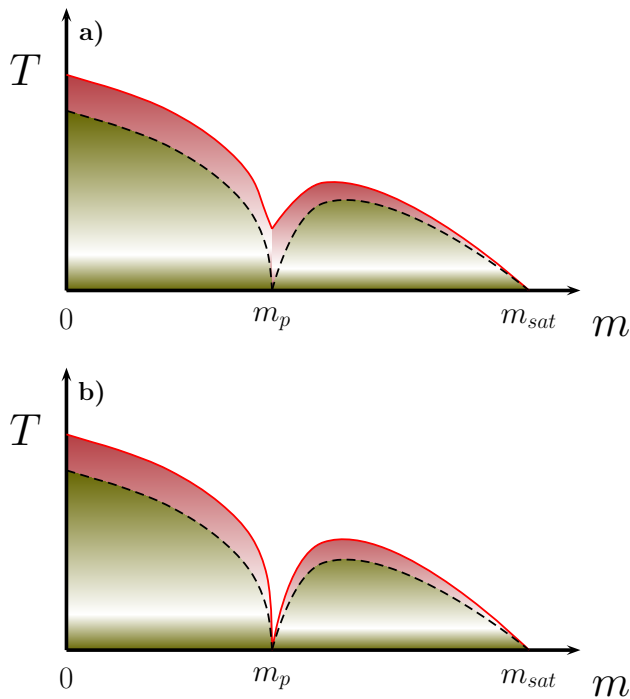}
\caption{(Color on-line) Possible scenarios for the KT and Ising transitions. Dashed black lines correspond to KT transitions
whereas solid red lines correspond to Ising transitions. In case (b) the $Z_2$ symmetry is restored at zero temperature by tunneling processes for values of the magnetization where the Berry phase term disappears.}
\label{fig:PH_diag}
\end{figure}

In our case, as the length of the planar component of the spin decreases with the magnetization or magnetic field $h$,
one expects also both critical temperatures to decrease with $h$. The situation becomes even more interesting in
the case where a magnetization plateau is present  (either $S-m$ integer or rational).
In this case, even at zero temperature the spin degrees of freedom have short range correlations implying $T_{KT} \to 0$.
The question whether $T_{Ising}$ also tends to zero for this value of the magnetization or remains non-zero is governed by the structure of the ground state of the quantum system and is discussed below.
In Fig.~\ref{fig:PH_diag} we plot the behavior of both critical temperatures as a function of magnetization for two different scenarios, in which the  chiral symmetry $Z_{2}$ is broken or not.

\section{Back to the quantum case: vorticity and Berry phase contributions}

In the calculation of the free-energy of  Section IV the vortex contribution to the action was not taken into account.
Indeed, the Matsubara decomposition of the fields $\phi_1$ and $\phi_2$ assumes periodicity in imaginary time and is only well defined
for smooth configurations, namely if the field configurations with vortices are excluded.
We turn to discuss the role of vortices below.

\subsection{Vorticity in the symmetric field }

For generic values of $S-m$,  destructive interference of different vortex configurations
\cite{TTH} due to the Berry phase term in Eq.(\ref{eq:sym_action}), 
vortex effects are indeed averaged out of the partition function.

In the particular case of integer $S-m$, 
the computation of the preceding section remains correct as long as the stiffness
of the fields is large enough to penalize vortex configurations.

When $S-m = {q \over p}$ is rational, vortices with vorticity $p$ can proliferate (larger $p$ weakens the effect). 
The first consequence of this scenario is the presence of a magnetization plateau, 
with degeneracy of the ground state given by $p$.
This is monitored by the behavior of the symmetric field
\ba
\phi_{s}=\frac12(\varphi_{1}+\varphi_{2})
\ea
which governs the physics of the total magnetization  of the system.
It is invariant under global rotations, thus a Goldstone mode of the system, and is conjugate to the fluctuation of the total magnetization $\Pi_1+\Pi_2$.
Its delocalization due to vortex proliferation translates into a locking of its conjugate
field at a fixed (quantized) value. This is nothing else than the presence of a magnetization plateau\cite{TTH,Lamas-Pujol_2010}.

\subsection{Vorticity in the anti-symmetric field }

The behavior of the anti-symmetric field
\ba
\phi_{a}=\frac12(\varphi_{1}-\varphi_{2})
\ea
is particular to the present model and more interesting here. Usually, such a field gets a mass term in the effective action. In the present case, the flatness of the potential obtained in Eq.~(\ref{eq:sym_action}) for the anti-symmetric field is the result of the continuous degeneracy of the classical ground states, not protected by any symmetry. Hence, it is not a Goldstone mode.
This field is conjugate to $\Pi_1-\Pi_2$,
directly linked to the relative spin angle $\alpha$ between sub-lattices. 
%

In this sense, the free energy shown in Fig.~(\ref{fig:numerical_Z_h0}) plays the role of a pseudo potential for $\phi_a$.
The presence of two minima in the pseudo potential is even more interesting,
since it allows for non-trivial tunneling processes, if allowed by the Berry phase term.
In weakly frustrated systems, the antisymmetric field is gapped due to the presence of
a mass term in the effective action. This term fixes the value of the field $\phi_a$ preventing vortex formation, and hence tunneling events. Then, in the weakly frustrated case, only the symmetric field $\phi_s$ may present a non-zero vorticity.
In the present case, due to the strong frustration, the antisymmetric field is not necessarily locked. The double minima potential allows processes where the vortices corresponding to $\phi_a$ can proliferate, disordering it and restoring 
the $Z_{2}$ symmetry. We discuss these kind of processes below.

\subsection{Zero temperature Ising transition}
In the preceding sections we have assumed that, at low enough temperatures, the emergent $Z_2$ symmetry is broken.
This is certainly the case if the tunneling between the two minima of the effective potential for $\phi_{a}$ is suppressed
by the Berry phase term, as it happens for generic values of $S-m$.

Let us now consider the case where $S-m$ is an integer. The Berry phase term can just be  dropped off from the action and the
computation of the partition function now allows the presence of processes where, in a localized region of the space,
the relative angle between the two sub-lattices goes from $0$ to $\pi$ and then goes back to $0$ again
when evolving in imaginary time. This process corresponds to the inclusion of tunneling between the two minima of the effective potential.
The inclusion of these kind of processes in the partition function is analogous to the traditional
low temperature expansion of the Ising model.
Indeed, the ``instanton-like'' excitations correspond to a small domain of ``$-$'' spins in a sea of otherwise fully ordered ``$+$'' spins in an effective three dimensional classical Ising model.
The energetic cost of a domain wall is proportional to the stiffness of the antisymmetric field $\phi_{a}$ which is in turn controlled by the couplings $J_1$ and $J_2$.

Reducing the value of the stiffness favors the proliferation of those instantons until a point
at which the system enters the disordered phase, where the $Z_2$ symmetry is restored.
Reducing the stiffness of $\phi_{a}$ can be easily achieved by approaching the limit $J_1 \to { J_2 \over 2}$.
Whether simply approaching this limit will be enough to restore the $Z_2$ symmetry or additional terms should be
included in the action is a question that goes beyond the scope of the present article but the possibility of a
zero temperature Ising transition is certainly an interesting issue that would deserve further analysis.

\section{Discussion and conclusions}

In this paper  we have studied the order-by-disorder selection in the $J_{1}-J_{2}$ Heisenberg model on the square lattice, in the presence of a magnetic field, by using a path integral approach. Quantum and thermal fluctuations select the collinear states from the largely degenerate manifold of classical ground states. The low energy effective theory of quantum fluctuations is written in terms of a symmetric field $\phi_{s}$, related to global magnetization, and an anti-symmetric field $\phi_{a}$ related to the spin imbalance between  sublattices. 

While the global magnetization is described by the symmetric field\cite{TTH}, we mainly focus on the $Z_2$ symmetry still present after order by disorder selection.
Let us recall that restoring the $Z_2$ symmetry implies disordering the field $\phi_{a}$ which in turn would be indicative of the following measurable effects:

{\it i)} The locking of the conjugate variable $\Pi_{a}$ to a (quantized) value which is not necessarily zero.
The conjugate variable to $\phi_{a}$ is nothing else than the difference of magnetization between sub-lattices 1 and 2,
also known as {\it spin imbalance}.
In this case not only the total magnetization would be locked to a special value (which reveals the disordering of
$\phi_{s}$ and the presence of a plateau in the magnetization curve) but also the difference of magnetization between
the two sub-lattices.

{\it ii)} The restoration of the rotational symmetry is also indicative of the formation of singlets in the system, triggering a transition
from a ground state pattern with a clear semiclassical interpretation of the spins (with no singlet formation in the ground state) to a more
quantum-mechanical and less degenerate ground-state. Such kind of plateau phases were dubbed classical and quantum plateaux by Hida and Affleck
in the study of a one-dimensional system \cite{Hida-Affleck_2005}.

{\it iii)} Another feature  which obviously is not unrelated to the effects mentioned above is
the factorization of the wave function into separable states.
Such kind of phenomena have been shown to occur in a large variety of quantum magnets which
are known to have OBD mechanism. 
In the presence of a magnetic field, and close to saturation, it has been 
shown that the wave function can be written as the tensor product of localized magnons
in a sea of polarized spins\cite{Honecker_magnons_2001}. 
In fact, the kagome model at the ${1 \over 3}$ plateau has a wave function which has a
large overlap with a test wave function consisting, again, in a tensor product of resonating
plaquettes \cite{honecker_kagome1/3}. 
Such kind of factorization also occurs in highly frustrated one dimensional
systems\cite{Matera-Lamas_2014,Plat_tube,Arlego-Rosales}.
In the zig-zag ladder it has been rigorously shown that a fully dimerized
state is the ground state of a family of Hamiltonians\cite{Matera-Lamas_2014}.  
 In fact, one could consider that the most known examples of factorized systems are
 the Majumdar-Gosh chain\cite{MG}, as well as the 2-dimensional Shastry-Sutherland 
 spin 1/2 system\cite{Shastry-Sutherland}.
 We should emphasize that in our approach the factorization prediction is a much more generic
 feature, as it is expected to be valid for any spin $S$.
In the case at hand, the factorization is expected to separate the two sublattices, with very little entanglement between sites belonging to each of them.

In systems with sometimes very high connectivity and a priori no simple reason for such a behavior of the ground state wave function, the factorization phenomenon and the locking of relative magnetizations mentioned above is to be considered as a smoking gun for a quantum mechanical manifestation of OBD.

\section{Acknowledgments}

PP acknowledge Xavier Plat for enlightening discussions. CAL, DCC and GLR are 
partially supported by PIP 0747 CONICET, PICT 1724 and PICT 2013-0009 ANPCyT.
PP is partially supported by the ``Agence Nationale de la Recherche''
under grant No. ANR 2010 BLANC 0406-0.
%
%
%
%
%
%
%
%
%
%
%
%
%
%
%
%
%
\bibliography{Referencias_Lamas.bib}

\end{document}